# Hybrid Fuzzy Control of Nonlinear Inverted Pendulum System

Abdulbasid Ismail Isa, Mukhtar Fatihu Hamza, and Mustapha Muhammad

*Department of Electrical and Electronics Engineering, Usmanu Danfodiyo, Sokoto, Nigeria*

*Department of Mechatronics Engineering, Bayero University, Kano, Nigeria*

eeabdulbasid@yahoo.com

mfhamza.mct@buk.edu.ng

mmuhammad.mcty@buk.edu,ng

*Abstract*—Complexity and nonlinear behaviors of inverted pendulum system make its control design a very challenging task. In this paper, a hybrid fuzzy adaptive control system using model reference approach is designed to control inverted-pendulum system. First, Lagrange model is used to develop the mathematical model of the system. Moreover, an adaptive fuzzy control system is developed to achieve position control and later simultaneous control of position and pendulum angle in the same control loop. The control algorithm is investigated to achieve control of objective of reference tracking, disturbance rejection and robustness to parameter variation. The performance of the proposed control scheme was compared with PID and LQR controllers, which numerical simulation show that the proposed control scheme provides high-performance dynamic characteristics and is robust with regard to parametric variations, disturbance and reference tracking.

*Keywords—Inverted pendulum, nonlinear model, Adaptive control, Fuzzy, LQR, PID*

## I. INTRODUCTION

Inverted pendulum system (IPS) is used extensively for teaching and research purposes, because of its nonlinear and unstable dynamic behaviours. These properties makes its control design a very challenging and a good benchmark of testing various control methods, ranging from conventional control to intelligent based control methods. Effective control design is very essential in an inverted pendulum system, for its application in humanoid walking control robots[1][2][3][4][5].

A lot of researchers have worked in addressing control problems associated of with the system. Hassan and Ling [6] proposed a PID controller for linear model of IPS to achieve independent control of cart position and angular position, the control design gives a very good steady state response. Jing *et al.* research cited as [7] "Design and Simulation of Fractional Order Controller for An Inverted Pendulum System" proposed a fractional order based PID controller for position and angular controls on linear model of IPS outperforms conventional PID control in terms of less overshoot and faster convergence. Yim *et al.* [8] applied PID control to stabilize IPS, the stability was achieved by careful tuning of controller parameters. Andrew [9] applied state feedback controller to achieve rod angle stabilization during strategic movement of the cart. Kumar *et al.*[10] Develops linear quadratic regulator control for IPS position and angular controls, in which the result showed LQR superiority as compared to state feedback and PID controllers. Sethi *et al* [11] compares model reference based Adaptive PID and fractional order for angular control of IPS, the result showed that fractional order based adaptive PID controller in terms of less overshoot, settling time and MSE

Moreover, a lot of researches on intelligent control are done to achieve stabilisation control of an inverted pendulum system. Lukuman and Magaji [3] compares GA tuned PID with conventional PID for position and angular control of IPS in which GA based PID outperform conventional PID in terms of settling time, rise time and overshoot. .Suresh *et al.*[12] developed fuzzy based model reference adaptive (MRAC) controller to control angular position of linear model of IPS, performance of the proposed controller was compared with conventional MRAC controllers, the result showed that (FMRAC) outperformed the conventional controllers in terms of delay time, rise time and settling time. Saifzul *et al.*[13] Propose self-erecting control on real system using T-S fuzzy with Adaptive Neuro-fuzzy inference designed based linear model using 16 fuzzy rules, the result showed the system's stability was guaranteed and experimental results proves effectiveness of the method. Fallahi *et al.* [14] demonstrated the effectiveness of Adaptive base NN PID controller in pendulum angle control of IPS over conventional PID controller in terms reference tracking and disturbance rejection. Mishra *et al.*[15] Compares fractional order PID and full order PID controllers optimized with GA for pendulum angle position of IPS, the performance of the controllers was assessed based minimization of the objective functions, and the result showed that fractional order based PID outperformed the full order one. The controllers studied above were applied on linear model of IPS; there performance might be limited around the operation points.

Manis *et al.* [16] presents a robust control strategy for IPS with uncertain disturbances using sliding mode control (SMC), integral based sliding mode (ISMC) controller was used to address tracking angle control of the system, the numerical simulation showed that ISMC is more robust to disturbance and more accurate as compared to conventional SMC super twisting SMC. Parsad *et al.* [17] applied LQR and PID to control nonlinear dynamical model of an inverted pendulum system, in which the simulation results shows that LQR has a comparative advantage over the PID based control system. Similarly, intelligent controllers were applied on nonlinear inverted pendulum system. For example, Ahmad *et al.*[18] Designed interval type-2 fuzzy logic controller (IT2-FLC) for nonlinear IPS and compares it with type-1 fuzzy PD controller for pendulum angle control. The simulation result showed that interval type-2 fuzzy logic controller has a good performance over wide range of uncertainties and external disturbances.

The goal of this paper was to design effective control scheme that can guarantee robustness external disturbance/parameter variation and reference tracking. This paper is organised as follows: Section II presents the structure and mathematical model of Inverted pendulum system and linearization at operating condition. System's control assessment is investigated in Section III, in order to buttress the control efficiency of the proposed scheme, linear quadratic regulator and PID control design is presented in Section IV so as to compare the performance of the prosed control method.

Numerical simulation results of an inverted-pendulum system under the possible occurrence of uncertainties are provided to demonstrate the robust control performance of the proposed control system in Section V. Result of numerical simulation will be presented in section VI, while Conclusions are drawn in Section VII.

## II. MATHEMATICAL MODEL AND LINERASATION OF INVERTED PENDULUM SYSTEM

The model of inverted pendulum cart was obtained using Lagrange method, which is one of the modelling methods used for dealing with complex systems.

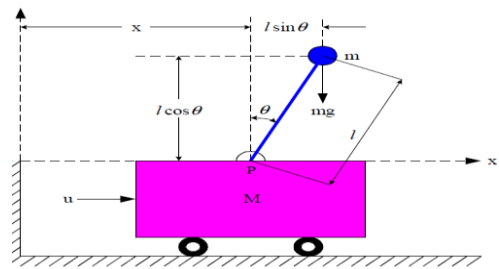

Fig. 1. Inverted Pendulum System[17]

The generalized coordinate of the system are $x(t)$ and $\theta(t)$. Therefore, Lagrange equation of the system is by:

$$L = \frac{1}{2}(M + m)\dot{x}^2(t) - ml\dot{x}(t)\dot{\theta}(t)\cos\theta(t) +$$
$$\frac{1}{2}ml^2\dot{\theta}^2(t) + mgl\cos\theta(t) + Fx(t) \qquad (1)$$

Dynamics of the system can be obtained by applying:

$$\frac{d}{dt}\left(\frac{\partial L}{\partial \dot{x}(t)}\right) - \frac{\partial L}{\partial x(t)} = 0 \qquad (2)$$

202

Therefore $x(t)$ dynamics can by written as:

$$\ddot{x}(t) = \frac{F + ml\ddot{\theta}(t)\cos\theta(t) - ml\dot{\theta}^2(t)\sin\theta(t)}{(M+m)} \qquad (3)$$

Similarly, $\theta(t)$ dynamics can be written by applying:

$$\frac{d}{dt}\left(\frac{\partial L}{\partial \dot{\theta}(t)}\right) - \frac{\partial L}{\partial \theta(t)} = 0 \qquad (4)$$

Therefore $\theta(t)$ dynamics can by written as:

$$\ddot{\theta}(t) = \frac{-mgl\sin\theta(t) + ml\ddot{x}(t)\cos\theta(t)}{ml^2} \qquad (5)$$

Equation (3) and (5) can be simplified as:

$$\begin{cases} \ddot{x}(t) = \dfrac{\left(F - ml\dot{\theta}^2(t)\sin\theta(t)\right)l - mgl\sin\theta(t)\cos\theta(t)}{-ml\dot{x}(t)\dot{\theta}(t)\sin\theta(t)\cos\theta(t)} \\[2pt] \qquad\qquad \dfrac{-ml\dot{x}(t)\dot{\theta}(t)\sin\theta(t)\cos\theta(t)}{(l(M+m) - ml\cos^2\theta(t))} \\[6pt] \ddot{\theta}(t) = \dfrac{\left(ml\cos\theta(t)\left(F - ml\dot{\theta}^2(t)\sin\theta(t)\right)\right)}{+\left(-mgl\sin\theta(t) - ml\dot{x}(t)\dot{\theta}(t)\sin\theta(t)\right)(M+m)} \\[2pt] \qquad\qquad \dfrac{}{((M+m)ml^2 - (ml)^2\cos^2\theta(t))} \end{cases} \qquad (6)$$

Our major concern is to keep the pendulum in the upright position around $\theta(t) = 0$, the linearization might be considered about this upright about equilibrium point. The linear model for the system around the upright equilibrium point is derived by simply linearization of the nonlinear system given in
$\theta(t) = 0$
$\sin\theta(t) \to \theta(t)$
$\cos\theta(t) \to 1$
$\dot{\theta}^2(t) \to 0$

Therefore equation (6) becomes

$$\begin{cases} \ddot{x}(t) = \dfrac{F - mg\theta(t)}{M} \\[6pt] \ddot{\theta}(t) = \dfrac{F - (M+m)g\theta(t)}{(Ml)} \end{cases}$$

The resultant linear model can be represented as:

$$\left.\begin{aligned} \dot{\mathbf{x}}(t) &= \mathbf{A}\mathbf{x}(t) + \mathbf{B}u(t) \\ \mathbf{y}(t) &= \mathbf{C}\mathbf{x}(t) + \mathbf{D}u(t) \end{aligned}\right\} \qquad (7)$$

Let

$$\mathbf{x}(t) = \begin{bmatrix} \theta(t) & \dot{\theta}(t) & x(t) & \dot{x}(t) \end{bmatrix}$$

Therefore, the system's equation can be written in a compact form as:

$$\mathbf{A} = \begin{bmatrix} 0 & 1 & 0 & 0 \\ \dfrac{-(M+m)}{Ml} & 0 & 0 & 0 \\ 0 & 0 & 0 & 1 \\ \dfrac{-gm}{M} & 0 & 0 & 0 \end{bmatrix}$$

$$\mathbf{B} = \begin{bmatrix} 0 & \dfrac{1}{Ml} & 0 & \dfrac{1}{M} \end{bmatrix}^T$$

$$\mathbf{C} = \begin{bmatrix} 1 & 0 & 0 & 0 \\ 0 & 1 & 0 & 0 \\ 0 & 0 & 1 & 0 \\ 0 & 0 & 0 & 1 \end{bmatrix}$$

$$\mathbf{D} = \begin{bmatrix} 0 & 0 & 0 & 0 \end{bmatrix}^T$$

Table I showed the inverted pendulum system parameters.

TABLE I.  SYSTEM PARAMETERS

| Parameter | Value |
|---|---|
| Acceleration due to gravity, $g$ | 9.8ms$^{-2}$ |
| Mass of the bob $m$ | 0.2kg |
| Mass of the Cart, $M$ | 1.2Kg |
| Length of pendulum, $l$ | 0.36m |

## III.  SYSTEM'S CONTROL ASSESMENT

The system's behaviour was asssessed based on it's parameters shown in Table I, and it was found to be state controrablle, observable and unstable.

## IV.  LINEAR QUADRATIC REGULATOR AND PID CONTROL DESIGN

The main objective of optimal control theory  is to determine control signals that will cause a process (plant) to satisfy some physical constraints and at the same time minimise a chosen performance criterion (performance index or cost function) given by:

$$J = \int_0^\infty \left( \mathbf{x}^T(t)\mathbf{Q}\mathbf{x}(t) + \mathbf{u}^T(t)\mathbf{R}\mathbf{u}(t) \right) dt \qquad (8)$$

Where $\mathbf{Q}$ and $\mathbf{R}$ are positive semi-definite and positive definite matrices respectively. The LQR gain vector $\mathbf{K}$ is given by:

$$\mathbf{K} = \mathbf{R}^{-1}\mathbf{B}^T\mathbf{P} \qquad (9)$$

Where $\mathbf{P}$ is a positive definite symmetric constant matrix obtained from the solution of matrix algebraic Riccati equation (ARE)

$$\mathbf{A}^T\mathbf{P} + \mathbf{P}\mathbf{A} - \mathbf{P}\mathbf{B}\mathbf{R}^{-1}\mathbf{B}^T\mathbf{P} + \mathbf{Q} = 0 \qquad (10)$$

The overall optimal control law is given by:

$$u(t) = -\mathbf{K}\mathbf{X}(t) + \mathbf{N}\mathbf{r}(t) \qquad (11)$$

Fig. 2 showed the SIMULINK model of LQR control applied to nonlinear inverted pendulum system.

The $-\mathbf{K}\mathbf{X}(t)$ component of equation (11) is used to stabilised system.

$\mathbf{N}$: Scaling matrix

$\mathbf{K}$: State feedback matrix

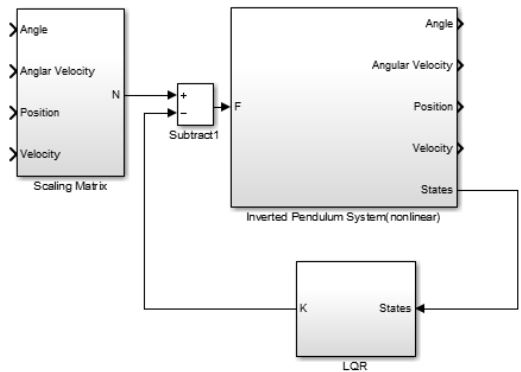

Fig. 2. LQR controlled System

While, This controller is commonly developed using three terms namely, proportional term, differential term and integral term combined together in a linear form[19]. The PID model in time domain is given as follows:

$$u(t) = K_p e(t) + K_i \int_0^t e(\tau)\, d\tau + K_d \frac{de(t)}{dt} \qquad (12)$$

Where $U(t)$ and $e(t)$ are control and error signals respectively. Similarly, $K_p, K_i$ and $K_d$ are proportional, integral and derivative constants respectively. The proportional term of PID reduces error due to disturbance; integral term eliminates steady-state error and the derivative term dampens the dynamic response, and hence improving the system stability.

Fig. 3 showed PID control of cart's position, while Fig. 4 depicts simultaneous control of PID of cart's position and pendulum angle of the system.

## V. HYBRID ADAPTIVE FUZZY CONTROLLER DESIGN

This section consist madel based adaptive control design and PI-D based fuzzy controller

### A. Adaptive Control Design

Generally, adaptive controller is a kind of a controller that can modify its behaviour in response to changes in the dynamics of the process and disturbance.

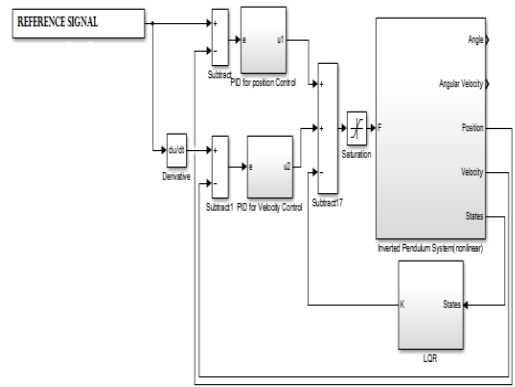

Fig. 3. PID control of cart's Position

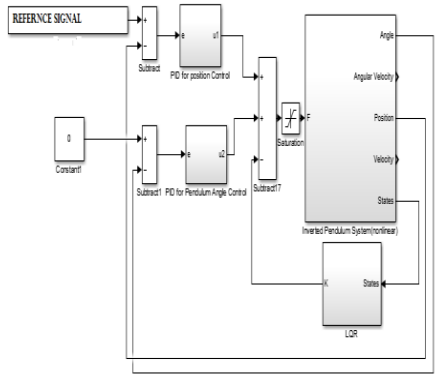

Fig. 4.    PID control of cart's and pendulum angle Positions

This research designed model reference based adaptive control, this control strategy addresses the control problem based on specific a given reference model.

Now let

$y(t)$: Plant output response

$y_m(t)$: Reference model output response for a given control action

The model error $e_m(t)$ is given by:

$$e_m(t) = y(t) - y_m(t) \qquad (13)$$

The adjustable parameter equation based on the given cost function $J$ is given by:

$$\frac{d\theta(t)}{dt} = -\gamma \frac{\partial J(t)}{\partial e(t)} \frac{\partial e_m(t)}{\partial \theta(t)} = -\gamma e(t) \frac{\partial e_m(t)}{\partial \theta(t)} \qquad (14)$$

The chosen control law is given by:

$$u(t) = \underbrace{K_p \lambda_1(t) + K_i \int_0^t \lambda_2(t)\, dt}_{Input1} + \underbrace{K_d \frac{d\lambda_3(t)}{dt}}_{input2} \qquad (15)$$

Where:

$$\lambda_1(t) = \theta_1(t) r(t) - \theta'(t) y(t)$$

$$\lambda_2(t) = \theta_2(t) r(t) - \theta'(t) y(t)$$

$$\lambda_3(t) = \theta_3(t) r(t) - \theta'(t) y(t)$$

In laplace form, the plant output can be written as:

$$y(s) = G(s) u(s) \qquad (16)$$

Therefore:

$$y(s) = \frac{G(s)\left(K_p \theta_1(s) r(s) + \frac{Ki}{s} \theta_2(s) r(s) + K_d s \theta_3(s) r(s)\right)}{(1 + 3G(s)\theta'(s))} \qquad (17)$$

We can apply equation(13), to obtain error model as:

$$e(s) = \frac{G(s)\left(K_p \theta_1(s) r(s) + \frac{Ki}{s} \theta_2(s) r(s) + K_d s \theta_3(s) r(s)\right)}{(1 + 3G(s)\theta'(s))}$$

$$G_m(s) r(s) \qquad (18)$$

Based on the above equations adjustable parameters can be written as:

$$\left. \begin{array}{l} \frac{d\theta_1(t)}{dt} \approx -\gamma_p e_m(t) y_m \\[4pt] \frac{d\theta_2(t)}{dt} \approx -\gamma_I e_m(t) y_m \\[4pt] \frac{d\theta_3(t)}{dt} \approx -\gamma_D e_m(t) y_m \\[4pt] \frac{d\theta'(t)}{dt} \approx -\gamma' e_m(t) y_m G_m(s) \end{array} \right\} \qquad (19)$$

### B. Fuzzy Control Design

Basic fuzzy controller consists of four functional basic blocks. These blocks include fuzzification, rule base, inference mechanism and defuzzificationA fuzzification is a conversion of crisp inputs into fuzzy membership values that are used in the rule base in order to execute the related rules so that an output can be generated, while inference mechanism represents the expert's decision making in interpreting and applying knowledge about how to control the plant. A defuzzification interface converts the conclusions of inference mechanism into the crisp control input for the process. A block diagram of fuzzy control system is shown in Figure 6. In this controller, variables are divided into input and output. This controller uses two input variables $(P + I)(k)$ and Change in error $\Delta e(k)$ one output variable $u(k)$.

The inputs to the FLC inputs has to be fuzzified. This is carried out using a membership function which maps the crisp inputs into a fuzzy set. In this

205

work triangular and trapezoidal membership functions are used. The linguistic variables used to describe the inputs $(P + I)(k)$ and $\Delta e(k)$ are Negative Big NB, Negative Medium NM, Negative Small NS, Zero Z, Positive Small PS, Positive Medium PM and Positive Big PB. Fig. 5 shows membership function diagram.

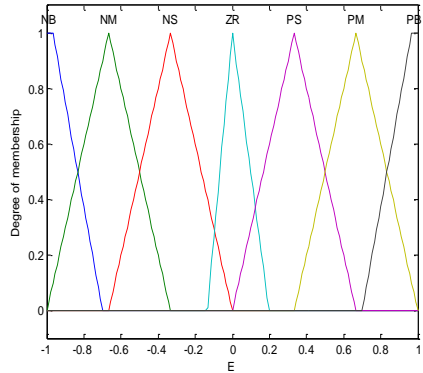

Fig. 5.    Membership Function

This research proposed co-opting the PID controller with intelligent adaptive based fuzzy controller so at as to improve the control efficiency and resolution. Simulink diagram of the prosed hybrid controller is shown Fig. 6.Therefore, the overall output of the proposed hybrid controller is given by:

$$U(.) = u(k) + K_p e(t) + K_i \int_0^t e(\tau)\, d\tau + K_d \frac{de(t)}{dt}$$
(20)

Where:

$$u(k) = \frac{\sum_{j=1}^{50} \mu_j(z_j) \boxtimes z_j}{\sum_{j=1}^{50} \mu_j(z_j)}$$

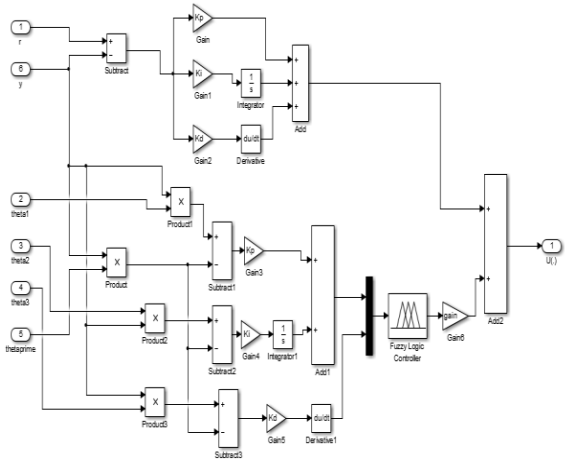

Fig. 6.    Simulink Block of Hybrid Fuzzy Controller

## VI.  RESULT AND DISCUSSION

The proposed control method was investigated via numerical simulations in MATLAB/Simulink. Controllers were design to achieve position tracking control of the cart and simultaneous control operation (position and pendulum angle control), the rejection efficiency and robustness to parametric variations of the prosed control will be assed. However, the weighting matrices are chosen as follows:

$$\mathbf{Q} = \begin{bmatrix} 1 & 0 & 0 & 0 \\ 0 & 9 & 0 & 0 \\ 0 & 0 & 230 & 0 \\ 0 & 0 & 0 & 180 \end{bmatrix} \text{ and } \mathbf{R} = 1.5$$

The LQR gain was obtained as:

$$\mathbf{K} = [2.0960 \quad -1.2221 \quad 12.3828 \quad 12.7813]$$

PID controller constants are shown in Table 2 as follows:

TABLE II.    PID CONSTANTS CART'S POSITION CONTROL

| Control variable | KP | Ki | KD |
|---|---|---|---|
| Position | 0.6 | 16 | 10 |
| Velocity | 10 | 8.9 | 0.009 |

TABLE III.    PID CONSTANTS CART'S POSITION AND PENDULUM ANGLE CONTROL

| Control variable | KP | Ki | KD |
|---|---|---|---|
| Angle | 6.9 | 0.009 | 1.4 |
| Position | 1 | 18 | 1 |

### A.  Cart Position Control

The simulation was carried out with a step input signal as the desired position of the cart. Fig. 7 showed system response of the developed position controllers at no disturbance, it can be seen that the three controllers performs satisfactorily in positioning the cart to the desired position. Table IV presents performance of the controllers, it can be seen that the proposed adaptive fuzzy controller outperforms LQR and PID controllers in terms of settling time, overshoot.

TABLE IV.    CART POSITION CONTROL AT NO disturbance

| Performance Indices | Hybrid Adaptive Fuzzy | LQR | PID |
|---|---|---|---|
| Settling Time (Seconds) | 6.1772 | 11.1301 | 11.5323 |
| Overshoot (%) | 0.6216 | 3.2985 | 18.0396 |
| Steady state error | 0 | 0.0319 | 0 |

| Performance Indices | Hybrid Adaptive Fuzzy | LQR | PID |
|---|---|---|---|
| Settling Time (Seconds) | 6.1501 | 12.6544 | 11.4962 |
| Overshoot (%) | 0.7667 | 3.3125 | 18.1397 |
| Steady state error | 0 | 0.0342 | 0 |

Similarly, the velocity response of the cart at no disturbance is shown Fig. 8, controllers efforts was in Table V presents the performance of the controllers. It can be seed that the proposed controller outperforms LQR and PID Controllers respectively, because its velocity becomes constant quickly with less overshoot.

The pendulum is disturbed by an external signal simulated in a form of random noise through its control input, while simulating the model. Fig. 9 showed the simulation result, which showed the proposed controller to be more robust to disturbance than the conventional controllers.

The effect of the pendulum disturbance was also investigated on the velocity response of the cart. Fig. 10 showed the in fact of controllers on velocity response in Table VI, which showed that the proposed controller is more robust to disturbance than the conventional controllers.

The performance of the proposed controller was also compared with that of the conventional controller due to parameteric variation at a step command by increaing the mass of the cart by 20% .Fig. 11 and shows the simulation result, which shows that the LQR controller fails to balance the stabilise the position of the while the proposed controller stablise the cart effectively. The velocity response and performance of the cart was in Fig. 12 and Table VIII respectively. It can be observed that the proposed showed superior rubusteness to paramere variation in comparison to conventional controllers. If fact LQR controlled system cannot produce enough control energy to giveout constant velocity.

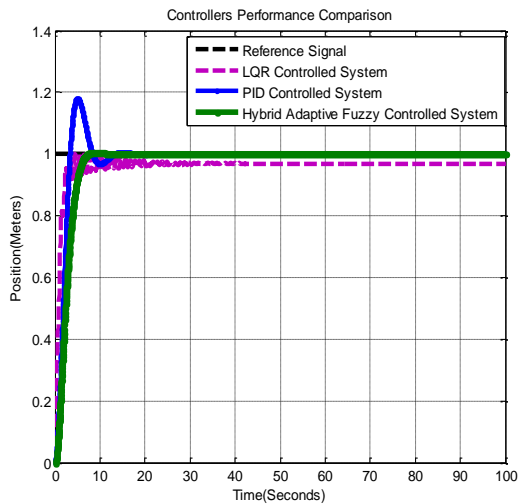

Fig. 7.   Carts's Position Response at no disturnace

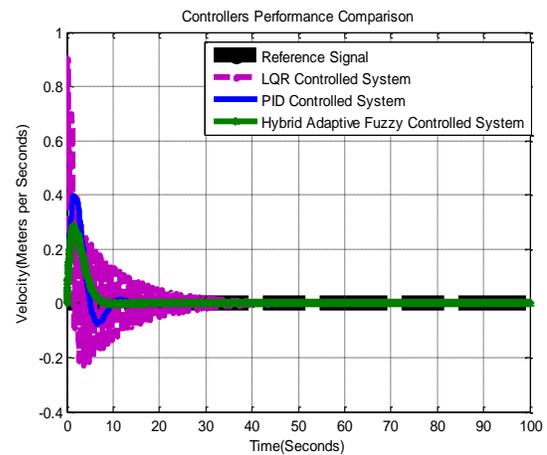

Fig. 8. Velocity Response of the Cart at no disturbance

TABLE V.       VELOCITY RESPONSE OF THE CART  AT NO DISTURBANCE

| Performance Indices | Hybrid Adaptive Fuzzy | LQR | PID |
|---|---|---|---|
| Settling Time (Seconds) | 8.1685 | 30.7334 | 13.1383 |
| Overshoot (%) | 0.3597 | 22.7591 | 7.4024 |

TABLE VI.       CART POSITION CONTROL DUE TO DISTURBANCE

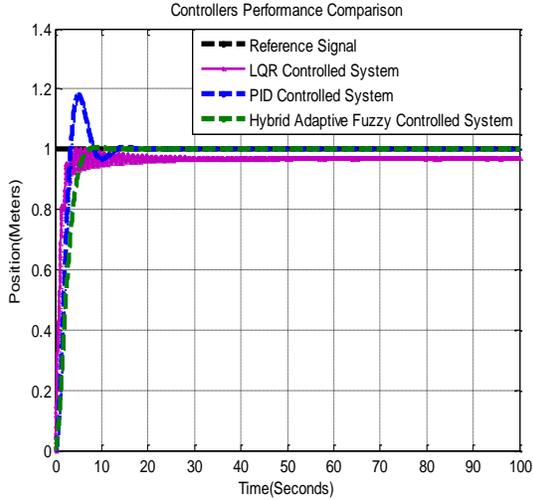

Fig. 9.        Carts's Position Response due to disturnace

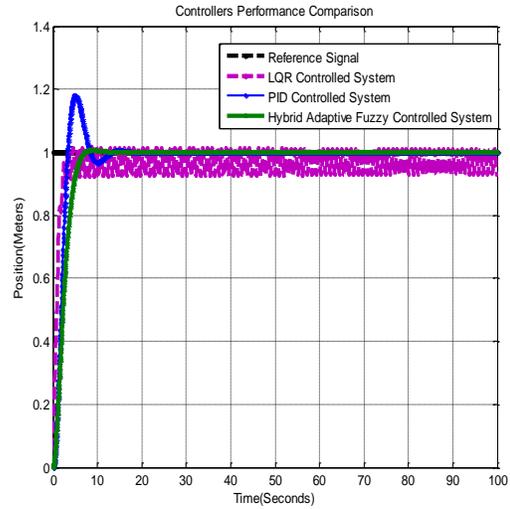

Fig. 11.        Cart's position Response due to parameter variation

Proceeding numerical showed the effectiveness of the proposed control over the conventional controllers in terms shorter settling time, less overshoot and robustness to disturbance/parameter variation. Fig. 13 and Fig. 14 showed the variation of settling time and overshoot for cart's position control in accordance with testing conditions and control methods.

TABLE VIII.    CART POSITION CONTROL DUE TO PARAMETER VARIATION

| Performance Indices | Hybrid Adaptive Fuzzy | LQR | PID |
|---|---|---|---|
| Settling Time (Seconds) | 6.1687 | 99.6906 | 11.5230 |
| Overshoot (%) | 0.6127 | 8.4322 | 18.1814 |
| Steady state error | 0 | Not stable | 0 |

Fig. 15 and Fig. 16 showed the variation of settling time and overshoot for cart's velocity response in accordance with testing conditions and control methods.

TABLE VII.    VELOCITY RESPONSE OF THE CART DUE TO DISTURBANCE

| Performance Indices | Hybrid Adaptive Fuzzy | LQR | PID |
|---|---|---|---|
| Settling Time (Seconds) | 12.3356 | 35.2848 | 12.9465 |
| Overshoot (%) | 7.8724 | 26.4599 | 9.2150 |

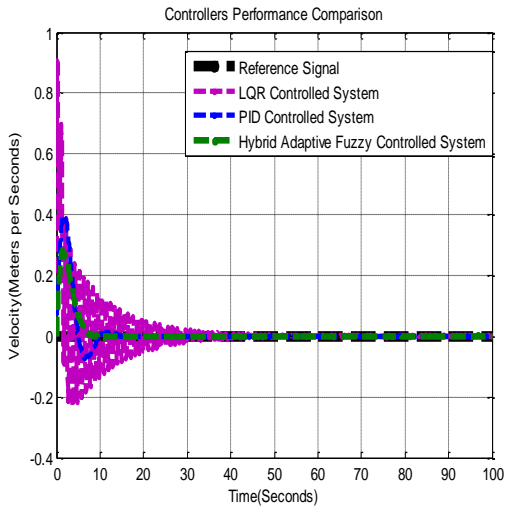

Fig. 10.    Velocity  Response due to disturnace

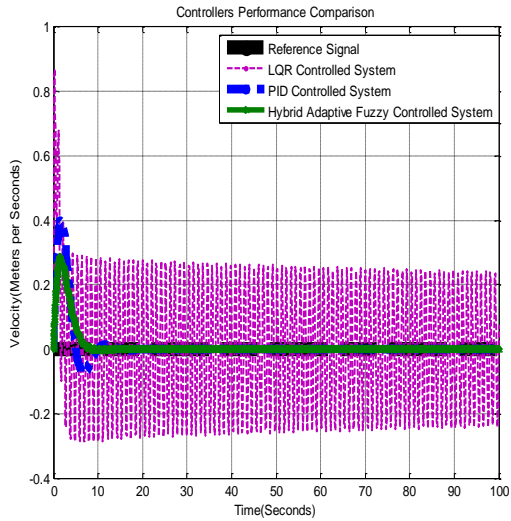

Fig. 12.    Velocity Response due to Parameter Variation

TABLE IX.    Velocity response of the cart Due PARAMETER Variation

| Performance Indices | Hybrid Adaptive Fuzzy | LQR | PID |
|---|---|---|---|
| Settling Time (Seconds) | 8.1870 | 99.9885 | 13.1070 |
| Overshoot (%) | 0.4014 | 39.4526 | 7.5498 |

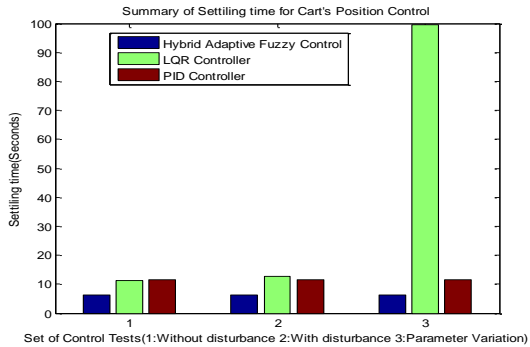

Fig. 13.    Settling time summary of Cart's position Control

## A.  Simulateneus Control

This section presents numerical simulation of cart's position and pendulum angle control in the same control loop. The developed controllers are expected in stabilise pendulum rod vertically up right at a particular reference cart position set at  0.3 meters, the simulation was carried out and system response was shown in Fig. 17 and Table IX showed the performance indices of the developed controllers.

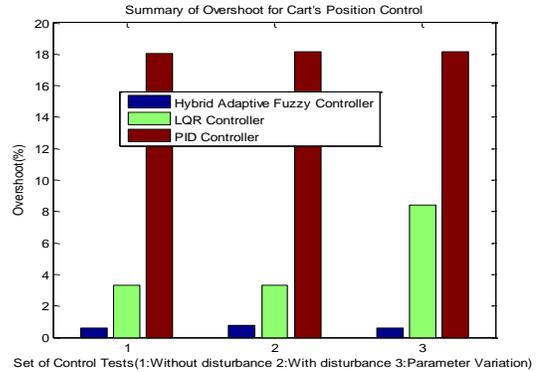

Fig. 14. Overshoot summary of Cart's Position Control

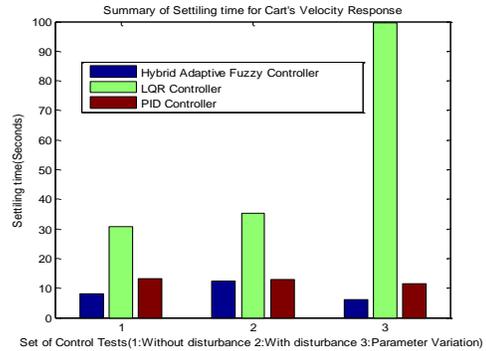

Fig. 15.   Settling time summary of Cart's Velocity Response

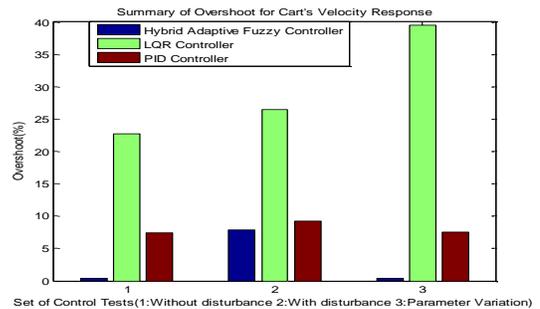

Fig. 16.   Overshoot summary of Cart's Velocity Response

It can be seen hybrid adaptive controller is more efficient in tracking command reference signal, for it shorter settling time and less overshoot.

TABLE X.        Cart Position Control At No disturbance

| Performance Indices | Hybrid Adaptive Fuzzy | LQR | PID |
|---|---|---|---|
| Settling Time (Seconds) | 7.7567 | 11.5004 | 8.7765 |
| Overshoot (%) | 3.1305 | 3.2251 | 29.8675 |
| Steady state error | 0 | 0.0094 | 0 |

Fig. 18. Pendulum angle position at no disturbance

TABLE XI. PENDULUM ANGLE CONTROL OF THE CART AT NO DISTURBANCE

| Performance Indices | Hybrid Adaptive Fuzzy | LQR | PID |
|---|---|---|---|
| Settling Time (Seconds) | 5.6920 | 40.5025 | 8.4096 |
| Overshoot (%) | 2.1444 | 15.9199 | 2.1795 |

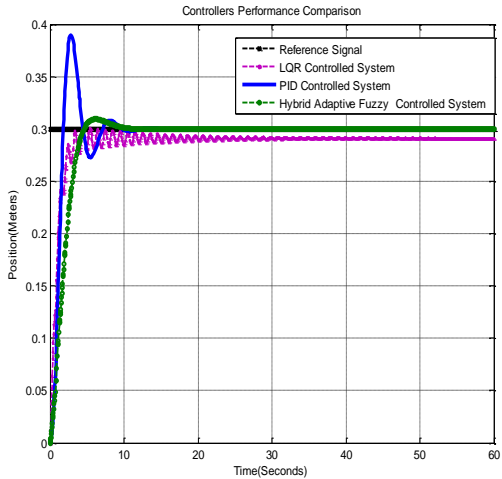

Fig. 17. Carts's Position Response at no disturnace

Pendulum angle response is shown in Fig. 18, it can be observed that hybrid adaptive fuzzy controller quickly stabilises the pendulum in the vertically upright position, followed by PID controller and later LQR. Table X summarised the performances indices of the developed controllers. Fig. 19 showed the cart's position response under the influence of disturbance. Table XI showed the performance of controllers in the presence of disturbance.

It wase observed that proposed hybdrid fuzzy controller is more rubust to disturbance as compared PID and LQR controllers.

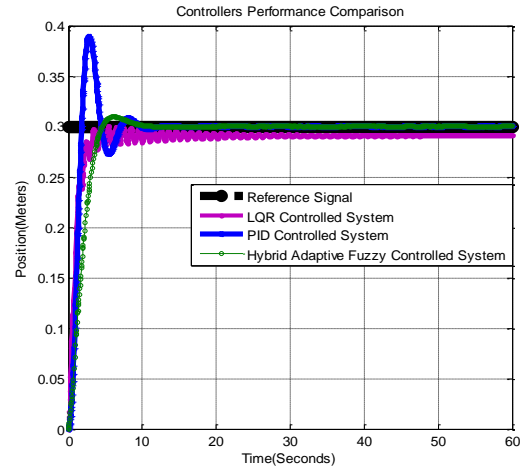

Fig. 19. Carts's Position Response due to disturnace

TABLE XII. CART POSITION CONTROL DUE TO DISTURBANCE

| Performance Indices | Hybrid Adaptive Fuzzy | LQR | PID |
|---|---|---|---|
| Settling Time (Seconds) | 7.7322 | 11.0007 | 8.7669 |
| Overshoot (%) | 3.1065 | 3.2343 | 29.8444 |
| Steady state error | 0 | 0.0097 | 0 |

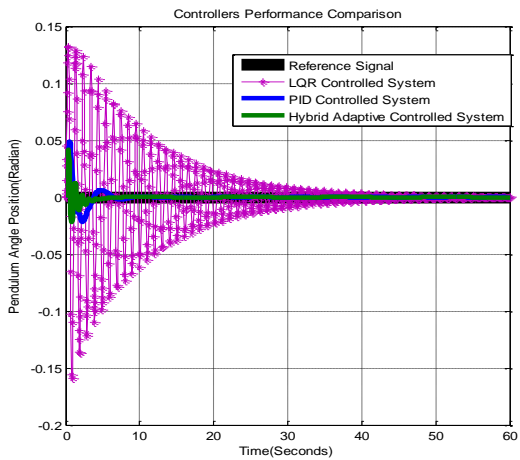

210

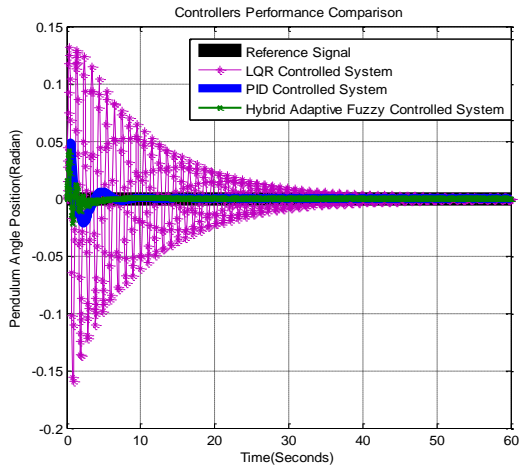

Fig. 20. Pendulum angle Position Response due to disturnace

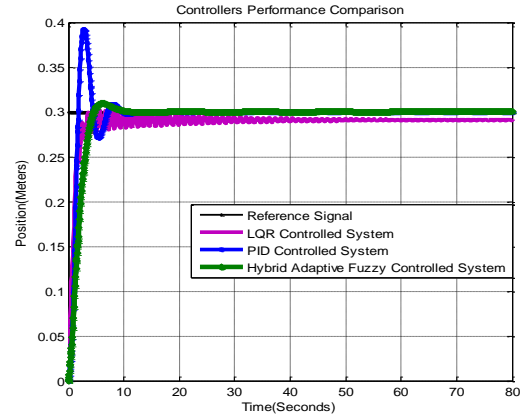

Fig. 21.   Cart's Position Response due to Parameter Variation

TABLE XIII.   PENDULUM ANGLE CONTROL OF THE CART DUE TO DISTURBANCE

| Performance Indices | Hybrid Adaptive Fuzzy | LQR | PID |
|---|---|---|---|
| Settling Time (Seconds) | 5.6787 | 44.0003 | 8.3552 |
| Overshoot (%) | 2.1424 | 15.8716 | 2.1714 |

TABLE XIV.   CART POSITION CONTROL DUE TO PARAMETER VARIATION

| Performance Indices | Hybrid Adaptive Fuzzy | LQR | PID |
|---|---|---|---|
| Settling Time (Seconds) | 7.7143 | 18.3265 | 8.8201 |
| Overshoot (%) | 3.1351 | 3.8197 | 30.8157 |
| Steady state error | 0 | 0.0086 | 0 |

Fig. 20 dipicts pendulum angle response in the presence disturbance, which  showed that hybrid adaptive fuzzy controller quickly stabilises the pendulum in the vertically upright position, followed by PID controller and later LQR.   Table XII summarised the performances indices of the developed controllers. Rubustness to parameter variation was assesd by increasing pendulum length by 5% and mass of the cart by15%.

Systems's response shown in Fig. 21 presents controllers performance,  which showed the rubustness the proposed control over the conventional control methods due to parametric variation as summarised Table XIII.

Pendulum angle response due to parameter  shown in Fig. 22, it is observed that hybrid adaptive fuzzy controller quickly stabilises the pendulum in the vertically upright position, followed by PID controller and later LQR as shown Table XIV.

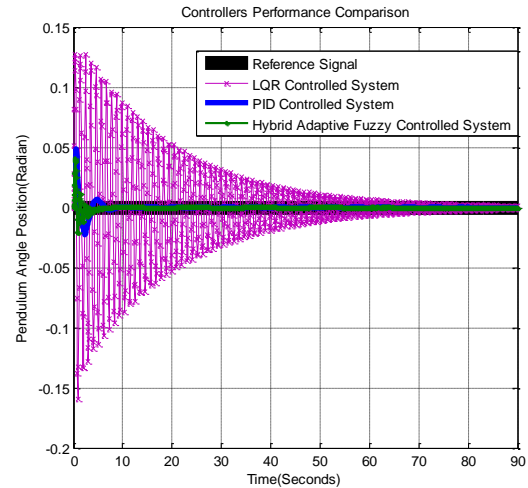

Fig. 22. Pendulum angle Position Response due to parameter variation

TABLE XV.   VELOCITY RESPONSE OF THE CART DUE TO PARAMETER VARIATION

211

| Performance Indices | Hybrid Adaptive Fuzzy | LQR | PID |
|---|---|---|---|
| Settling Time (Seconds) | 5.8317 | 85.2165 | 8.4312 |
| Overshoot (%) | 2.0910 | 16.0875 | 2.2572 |

The numerical simulation showed the effectiveness of the proposed control over the conventional controllers in terms shorter settling time, less overshoot and robustness to disturbance/parameter variation. Fig. 23 and Fig. 24 showed the variation of settling time and overshoot for cart's position control in accordance with testing conditions and control methods.

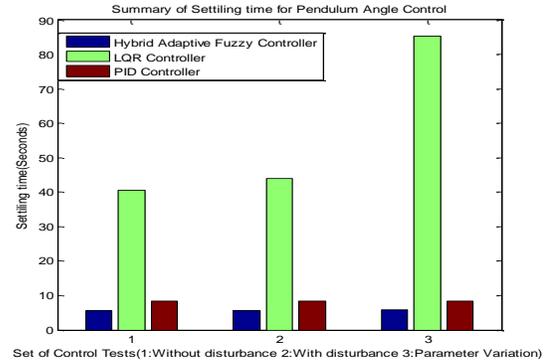

Fig. 25.        Settling time summary of Pendulum Angle position Control

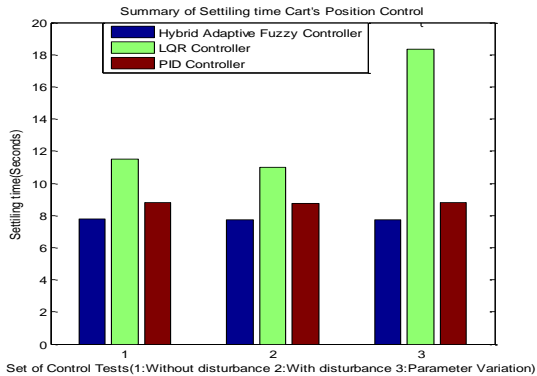

Fig. 23.        Settling time summary of Cart's position Control

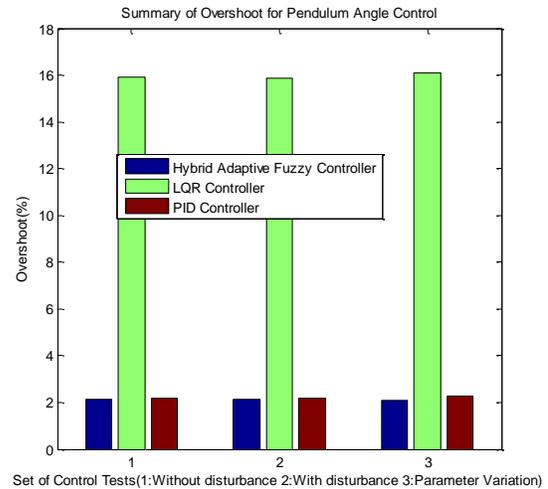

Fig. 26.        Overshoot summary of Pendulum Angle position Control

Fig. 25 and Fig. 26 showed the variation of settling time and overshoot for pendulum angle position control in accordance with testing conditions and control methods.

VII.    CONCLUSION

The mathematical model of  inverted pendulum system was derived and system control behaviour was also assessed. Adaptive based fuzzy controller was applied on nonlinear model of test system so at invesgitate the control efficiency of proposed method. The performance of  the control scheme was invesigated through numerical simulations with MATLAB, the control system was tested at various testing conditions; the result was compared with convetional controllers natabley PID and LQR controllers.The first section presents numerical simulation cart position and performace indices of the proposed controllers. Intelligent adaptive Fuzzy controller is faster than PID controller, because it takes 54% PID's and 56% of LQR's settling times to tracks the desired position at no disturbance. Also the overshoot value of the proposed controller is 3.45%,18.8% of PID and LQR controllers respectively at the same testing condition. disturbance was applied to the system so as to asses Disturbance rejection capacity, fuzzy based control showed rubustness over the conventional controllers, because it takes 53% PID's and 49% of LQR's settling times to tracks the desired position in the presence of disturbance,similarly, overshoot value of the proposed controller is 4.2%,23.1% of PID and LQR controllers respectively.Rubustness to parameter variation was also invegitated, which proves the propsed control more rubust to parameter variation for its adaptive nature.

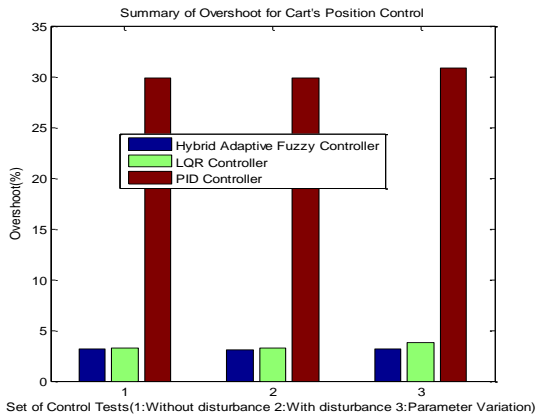

Fig. 24.        Overshoot summary for Cart's position Control

Second section the simulation presents simutaneus control cart's position and pendulum angle position in same control loop. Performance of the developed controllers was compared at different testing conditions as shown in section V. Fuzzy adaptive based control at no disturbance condition proves to be fast in response because it takes 67%, 83% of LQR's and PID's settling time, similarly, overshoots value is less as compared to PID and LQR controllers. Fuzzy controller takes 97.07%,10.5% of overshoot value of LQR and PID controllers.

Disturbance rejection capacity was also assesed, it was shown that the proposed control method is more efficient in terms of settling time and overshoot. Fuzzy controller takes 73% of LQR and 88.2% of PID settling time's to settle to it's final value.

Conclusively, the proposed control scheme is more accurate in reference tracking, faster in response and more robust to disturbance.

REFERENCES

[1]    A. I. Isa, M. F. Hamza, and M. Muhammad, "Hybrid Fuzzy Control of Nonlinear Inverted Pendulum System."

[2]    N. Magaji, M. F. Hamza, and A. Dan-Isa, "Comparison of GA and LQR tuning of static VAR compensator for damping oscillations," *international Journal Of Advances In Engineering & Technology,* vol. 2, p. 594, 2012.

[3]    A. A. H. Chiroma, A. Khan, M. F. Hamza, A. B. Dauda, M. Nadeem, S. Asadullah*, et al.*, "Utilizing Modular Neural Network for Prediction of Possible Emergencies Locations within point of Interest of Hajj Pilgrimage," *Modern Applied Science,* vol. 8, 2014.

[4]    M. F. Hamza, H. J. Yap, and I. A. Choudhury, "Genetic algorithm and particle swarm optimization based cascade interval type 2 fuzzy PD controller for rotary inverted pendulum system," *Mathematical Problems in Engineering,* vol. 2015, 2015.

[5]    A. Abubakar, T. Mantoro, S. Moedjiono, H. Chiroma, A. Waqas, M. F. Hamza*, et al.*, "A Support Vector Machine Classification of Computational Capabilities of 3D Map on Mobile Device for Navigation Aid," *International Journal of Interactive Mobile Technologies (iJIM),* vol. 10, pp. 4-10, 2016.

[6]    A. Y. Zimit, H. J. Yap, M. F. Hamza, I. Siradjuddin, B. Hendrik, and T. Herawan, "Modelling and Experimental Analysis Two-Wheeled Self Balance Robot Using PID Controller," in *International Conference on Computational Science and Its Applications*, 2018, pp. 683-698.

[8]    Y. Zhang and J. Wang, "Modelling , Controller Design and Implementation for Spherical Double Inverted Pendulum System," in *11th World Congress on Intellegent Control and Automation*, 2014, pp. 2967–2972.

[9]    L. Wai, Rong-jong, Chang, "Adaptive Stabilizing and Tracking Control for a Nonlinear Inverted-Pendulum System via," *IEEE Trans. Ind. Electron.*, vol. 53, no. 2, pp. 674–692, 2006.

[10]   N. M. Yusuf Lukman A, "GA-PID Controller for Position of Inverted Pendulum," in *6th IEEE International Conference on Adaptive Science and Technology*, 2014, pp. 1–21.

[11]   M. T.-P. G. Ronquillo-Lomeli, G.J Rios-Moreno, A.Gomez-Espinosa, L.A Moreles-Hernandez, "Nonlinear identification of inverted pendulum system using Volterra polynomials," *Mech. Based Des. Struct. Mach.*, vol. 44, no. 1, pp. 5–15, 2016.

[12] G. Tirian, O. Pro, I. Filip, and C. Ra, "Inverted pendulum controlled through fuzzy logic," in *10th Jubilee IEEE International Symposium on Applied Computational Intelligence and Informatics*, 2015, pp. 85–90.

[13] N. Hasan, L. I. Ling, Y. De-cheng, and J. Yuan-wei, "Modeling and Simulation of the Inverted Pendulum Control System," in *27th Chinese Control and Decision Conference*, 2015, pp. 548–552.

[14] J. Shuhua, M. Li, C. Li, and Wang, "Design and Simulation of Fractional Order Controller for An Inverted Pendulum System," in *IEEE International Conference on Manipulation, Manufacturing and Measurement on the Nanoscale*, 2017, pp. 6–9.

[15] Y. Y. Lim, C. L. Hoo, Y. Myan, and F. Wong, "Stabilising an Inverted Pendulum with PID Controller," in *Eureca2017*, 2017, pp. 1–14.

[16] Stimac K Andrew, "Standup and Stabilization of the Inverted Pendulum," Massachusetts Institute of Technology, 1999.

[17] R. Kumar, Pankaj Mukherjee, "Modelling and Controller Design of Inverted Pendulum," *Int. J. Adavanced Res. Comput. Eng. Technol.*, vol. 2, no. 1, pp. 200–206, 2013.

[18] D. Sethi, J. Kumar, and R. Khanna, "Design of Fractional Order MRAPIDC for Inverted Pendulum System," *Indian J. Sci. Technol.*, vol. 10, no. August, 2017.

[19] B. P. Suresh, S. P. Jadhav, P. Khalane, S. B. Pingale, P. Jadhav, and V. P. Khalane, "Design Of Fuzzy Model Reference Adaptive Controller For Inverted Pendulum," in *International Conference on Information Processing*, 2015, pp. 790–794.

[20] A. A. Saifizul, Z. Zainon, N. A. A. Osman, C. A. Azlan, and U. F. S. U. Ibrahim, "Intellenget Control for Self-erecting Inverted Pendulum Via Adaptive Neuro-Fuzzy Inference System," *Am. J. Appl. Sci.*, vol. 3, no. 4, pp. 1795–1802, 2006.

[21] M. Fallahi, "Adaptive Control of an Inverted Pendulum Using Adaptive PID Neural Network," in *International Conference on Signal Processing Systems*, 2009, pp. 589–593.

[22] S. K. Mishra and D. Chandra, "Stabilization of Inverted Cart-Pendulum System Using PI $\lambda$ D $\mu$ Controller : A Frequency-Domain Approach," *Chinese J. Eng.*, vol. 2013, pp. 1–7, 2013.

[23] M. Yadav, A. K. Gupta, B. Pratap, and S. Saini, "Robust Control Design for Inverted Pendulum System with Uncertain Disturbances," in *First IEEE International Conference on Power Electronics. Intelligent Control and Energy Systems*, 2016, pp. 1–6.

[24] L. B. Prasad, "Optimal Control of Nonlinear Inverted Pendulum System Using PID Controller and LQR : Performance Analysis Abstract :," *Int. J. Autom. Comput.*, vol. 11, no. December, pp. 661–670, 2014.

[25] A. M. El-nagar, M. El-bardini, and N. M. El-rabaie, "Intelligent control for nonlinear inverted pendulum based on interval type-2 fuzzy PD controller," *Alexandria Eng. J.*, vol. 53, no. 1, pp. 23–32, 2014.

[26] A. I. I. M.F Hamza, "Effect Of Sampling Time On PID Controller Design For A Heat Exchanger System," in *6th IEEE International Conference on Adaptive Science and Technology*, 2014.